\documentclass[letterpaper]{article} 
\usepackage{aaai2026}  
\usepackage{times}  
\usepackage{helvet}  
\usepackage{courier}  
\usepackage[hyphens]{url}  
\usepackage{graphicx} 
\usepackage{natbib}  
\usepackage{caption} 
\usepackage{algorithm}
\usepackage{algorithmic}

\usepackage{newfloat}
\usepackage{listings}
\DeclareCaptionStyle{ruled}{labelfont=normalfont,labelsep=colon,strut=off} 
\floatstyle{ruled}
\newfloat{listing}{tb}{lst}{}
\floatname{listing}{Listing}
\title{T-REX: Transformer-Based Category Sequence Generation for Grocery Basket Recommendation}

\author{
	Soroush Mokhtari\textsuperscript{\rm 1}, 
	Muhammad Tayyab Asif\textsuperscript{\rm 2},
	Sergiy Zubatiy\textsuperscript{\rm 1}, 
}
\affiliations{
	 \textsuperscript{\rm 1}Amazon WorldWide Grocery\\
	 \textsuperscript{\rm 2}Fulfilled By Amazon\\
	
	sorous@amazon.com,
	asifmt@amazon.com,
	zubatiy@amazon.com
}

\begin{document}

\maketitle

\begin{abstract}
Online grocery shopping presents unique challenges for sequential recommendations due to repetitive purchase patterns and complex item relationships within the baskets. Unlike traditional e-commerce, grocery recommendations must capture both complementary item associations and temporal dependencies across shopping sessions. To address these challenges in Amazon's online grocery business, we propose T-REX, a novel transformer architecture that generates personalized category-level suggestions by learning both short-term basket dependencies and long-term user preferences. Our approach introduces three key innovations: (1) an efficient sampling strategy utilizing dynamic sequence splitting for sparse shopping patterns, (2) an adaptive positional encoding scheme for temporal patterns, and (3) a category-level modeling approach that reduces dimensionality while maintaining recommendation quality. Although masked language modeling techniques like BERT4Rec excel at capturing item relations, they prove less suitable for next basket generation due to information leakage issues. In contrast, T-REX's causal masking approach better aligns with the sequential nature of basket generation, enabling more accurate next-basket predictions. Experiments on large-scale grocery offline data and online A/B tests show significant improvement over existing systems.
\end{abstract}

\section{Introduction and background}
Personalization is one of the core features of the modern e-commerce. While retail shopping typically involves one-off purchases (appliances, furniture, clothing), grocery shopping is characterized by repeated purchases and related items within shopping sessions. This distinction makes grocery recommendations fundamentally different from typical e-commerce scenarios - instead of suggesting exploratory purchases, we need to generate predictions for entire shopping baskets that align with established purchase patterns.

Previous works have shown that recommendation performance significantly depends on dataset characteristics \cite{shao2022A, li2023A}. Traditional personalization approaches such as customer-similarity and latent-affinity models excel at exploratory interactions but struggle with grocery's structured, repetitive nature and underperform compared to simple baselines like personal top-frequency (P-Top) \cite{li2023A}, which is currently deployed in Amazon's grocery personalization.

Recently, Next Basket Recommendation (NBR) has gained considerable attention \cite{Faggioli2020Recency, Liu2016A, Hu2019Set, liu2020Basket, shao2022A}, with focus on generative approaches that apply sequential neural architectures \cite{Liu2016A, Sun2019Bert, Moreira2021Transformer, Li2023Text, kang2018Self}. In this regard, transformer networks, which have revolutionized natural language processing through models like BERT \cite{devlin2018Bert} and GPT \cite{radford2018Improving}, are particularly promising due to their ability to learn complex sequential patterns and generate contextually appropriate outputs.

In this work, we propose T-REX, a transformer architecture specifically designed to generate personalized recommendations for grocery categories. Our approach differs from previous transformer recommenders in three key aspects: First, we introduce a dynamic sequence splitting method that enables efficient training while preserving shopping patterns across different time scales. Second, we develop an adaptive positional encoding scheme that better captures the irregular temporal patterns inherent in grocery shopping. Third, we implement category-level modeling which not only addresses computational challenges, it also aligns with how customers naturally organize their shopping lists.

Results from A/B tests in production environments show that T-REX significantly improves upon existing systems at Amazon Grocery Businesses. To further understand the efficacy of our recommendations, we introduce a novel rank-matching evaluation approach that quantifies how well our generated recommendations align with shopping patterns. This metric is particularly important for grocery shopping, where the order of category presentations affects basket building efficiency.

The rest of the paper is structured as follows. Section 2 formalizes the next basket prediction problem. Section 3 details our transformer architecture and its key innovations. Sections 4 and 5 describe our dataset and evaluation framework. Section 6 presents experimental results.

\section{Problem formal statement}
For grocery domain, the task of next-basket prediction involves generating complete shopping baskets from customers' purchase histories. For a customer $u \in \mathcal{U}$, we represent their purchase history as a temporally ordered sequence $S^{(u)}=(S_1^{(u)} ,\dots,S_t^{(u)})$, where $S_t^{(u)}$ indicates the shopping session at time $t$. Given the collective purchase history $S=\cup_{u \in \mathcal{U}} S^{(u)}$, our model generates probability distribution $P(S_{t+1}^{(u)}|S)$ over the possible next baskets.

Each basket $S_t^{(u)}$ contains category-level items $v \in \mathcal{V}$, represented as $S_t^{(u)}={v_t^{(u)} | u\in \mathcal{U}, v \in \mathcal{V}}$. While purchase histories $S^{(u)}_{t}$ follow temporal order (denoted by $(.)$), individual shopping sessions $S_t^{(u)}$ are unordered sets (denoted by ${.}$). This distinction is important: unlike physical stores where the layout may influence product selection order, online grocery shopping shows no consistent temporal patterns within baskets.

To make this generation task tractable while maintaining practical utility, we map the original 29K products to 35 categories using Amazon's Grocery taxonomy. This transformation addresses several key challenges:

\begin{itemize}
	\item Basket size variability becomes more manageable as category-level baskets have natural upper bounds (maximum 35 categories vs. arbitrary product counts)
	\item Cross-category dependencies are directly modeled instead of sparse product-to-product relationships
	\item Temporal patterns become more robust as category-level preferences are more stable than product-specific choices
	\item Seasonal variations are captured more reliably through category-level trends rather than volatile individual product fluctuations
	\item Computational complexity reduces dramatically from 29K to 35 possible predictions, making real-time inference feasible at scale
\end{itemize}

This transformation aligns with how customers typically plan their shopping - thinking in terms of categories (e.g., "dairy," "produce") rather than specific products.

\section{Transformer model}
While transformers have revolutionized sequence modeling, their application to grocery basket generation presents unique challenges. The original transformer \cite{vaswani2017Attention} proposed an encoder-decoder structure for machine translation. Although variants like BERT \cite{devlin2018Bert} and GPT \cite{radford2018Improving} have shown success with single-module architectures, we find the the encoder-decoder structure is better suited for generating grocery recommendations for several key reasons.

BERT4REC \cite{Sun2019Bert}, through masked language modeling with Cloze objectives \cite{Kolen2001A}, effectively captures item relationships. However, it faces limitations in basket building. We found two critical issues with the bidirectional approach. First, when predicting future purchases, the model's prediction error approached zero unless we masked all instances of the target product in the input sequence, indicating that the bidirectional model could "peek" at future occurrences of the same product—creating an unrealistic advantage not available in real-world recommendation scenarios. Second, the masking strategy poses fundamental challenges for the generation task: during inference, we can only utilize predictions for the last masked token to generate recommendations, as other masked positions don't directly contribute to the generation task. While these additional masks help learn product relationships, they fundamentally misalign with the sequential nature of basket generation.

Our T-REX architecture employs distinct encoder and decoder roles that prove essential for grocery shopping, where both complementary item relationships and temporal dependencies drive purchase decisions:
\begin{itemize}
	\item Encoder: Learns generalizable basket composition patterns across the customer base while remaining flexible to incorporate additional signals such as customer impressions, search queries, and browsing behavior
	\item Decoder: Generates personalized predictions by leveraging both population-level patterns (via cross-attention to encoder outputs) and individual shopping histories
\end{itemize}

This architectural separation naturally suits grocery basket generation, where we need to balance population-level patterns with individual preferences. The encoded context is seamlessly communicated to the decoder through cross-attention mechanisms, enabling more nuanced and personalized recommendations. Our implementation introduces several modifications to improve generation quality and efficiency.

\subsection{Sampling and batching input data}
\begin{figure*}
	\centering
	\includegraphics[width=0.8\textwidth]{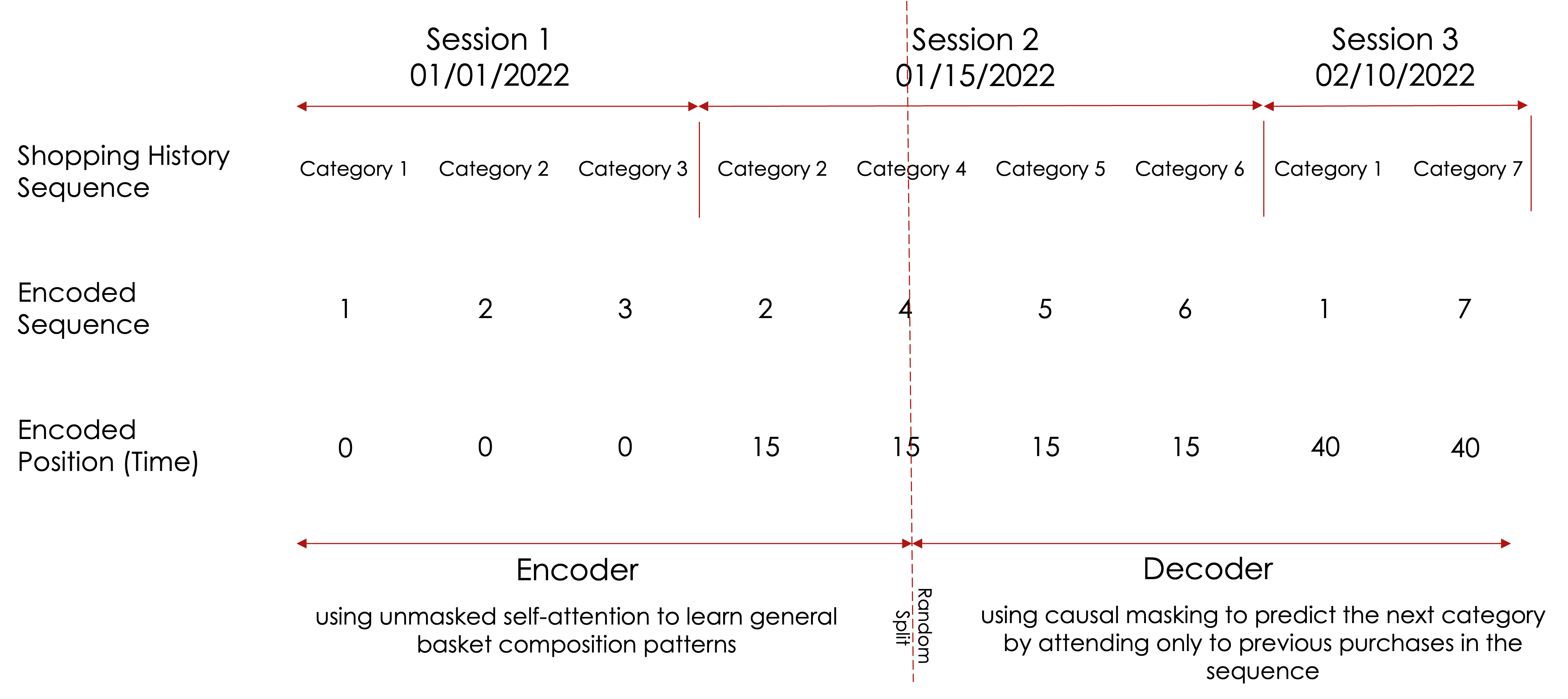}
	\caption{Illustration of dynamic sequence splitting and corresponding attention mechanisms in T-REX. The example shows a synthetic purchase sequence used for demonstration purposes only.}
	\label{fig:dynamic_split}
\end{figure*}

We generate training instances through dynamic sequence splitting, where each customer history $S^u$ yields multiple training samples. For each training instance, we select a pivot purchase $v_p^u \in S_t^u$ that divides the sequence into encoder input $S_{enc}^u$ and decoder target $S_{dec}^u$. The pivot selection follows a time-dependent strategy. First, we randomly sample a timestamp $t_p$ within each customer's shopping history timeframe. The purchase event closest to this timestamp becomes the pivot $v_p^u$. All purchases before $t_p$ form the encoder input $S_{enc}^u$, while the subsequent $n$ purchases after $t_p$ (where $n$ is a hyperparameter) form the decoder target $S_{dec}^u$. The dynamic splitting is illustrated in Figure \ref{fig:dynamic_split}.

This dynamic splitting approach offers several key advantages:
\begin{itemize}
	\item Creates diverse training contexts by sampling different temporal points.
	\item Improves generalization for new customers by learning from partial histories.
	\item Better handles varying history lengths through flexible pivot selection.
	\item Maintains temporal consistency in the split sequences.
\end{itemize}

While alternative approaches such as session-based splitting (using '$<$EOS$>$' tokens) or intra-basket shuffling were explored, the dynamic splitting strategy proved more effective, particularly for real-world deployment scenarios. Unlike session-based splitting, our approach enables the model to generate recommendations mid-session by treating the current basket as a partial sequence, making it particularly suitable for real-time recommendation generation. This capability is crucial for practical applications where customers expect relevant suggestions as they build their baskets.

For efficient batch processing, we stack these split sequences with padding token $v_{pad}$ to uniform length $n$.

\subsection{Product and positional embeddings}
Each product in the customer purchase history is represented as one token, mapped directly to its UPH category without incorporating additional product attributes. We represent these categories in the form of an embedding matrix $O \in R^{|\mathcal{V}| \times d}$, where $d$ is the latent dimensionality, yielding sequence embeddings $E^u \in R^{n \times d}$ for each customer. Unlike recurrent architectures (GRUs, LSTMs), transformers require explicit position information through embeddings $P^u \in R^{n \times d}$, added to category embeddings as $E^u \leftarrow E^u+P^u$.

Grocery shopping's irregular temporal patterns create unique challenges for position encoding. While NLP applications assume sequential positions, shopping sessions can be days or weeks apart. We explored several approaches to capture these temporal relationships:

\begin{itemize}
	\item Standard sinusoidal encodings \cite{vaswani2017Attention} using absolute positions, which proved inadequate for sparse temporal patterns.
	\item Learnable absolute position embeddings, which struggled with varying time intervals.
	\item Weekly position encoding to reduce sparsity, trading temporal precision for better generalization.
	\item Sigmoid-transformed absolute positions to preserve local temporal relationships while dampening long-term effects.
	\item Sequential positions relative to pivot \cite{Sun2019Bert, kang2018Self}, which lost important temporal information.
	\item Normalized positions relative to pivot purchase (our chosen approach), where positions are represented as days before/after the pivot purchase.
\end{itemize}

\subsection{Self-attention and multi-head generation}
Our generation process utilizes batched embeddings ($E \in R^{b \times n \times d}$) in self-attention layers to capture purchase patterns. For each position, the model generates three projections: key $K = EW^K$, query $Q=EW^Q$, and value $V=EW^V$. Here $\{W^{i}\}_{i = \{K, Q, V \}}$ contains the latent relationships between various sets of grocery items within the baskets $E$, learnt via back-propagation. To this end, in the forward pass, we compute the standard attention array:

\begin{equation}
	H(E)= softmax((QK^T)/\sqrt{d})V
\end{equation}

where scaling by $\sqrt{d}$ maintains stable gradients during training. To enrich different aspects of shopping patterns (from large volume of historical purchases), we employ parallel attention heads, concatenating their outputs:

\begin{equation}
	M_H (E)=Concat(H_1, \dots ,H_h ) W^O
\end{equation}

The encoder processes all historical purchases without masking to learn general basket composition patterns. The decoder, through causal masking, generates predictions by relying only on previous purchases. This way it is able to learn the temporal causal relations. Each layer's output passes through position-wise feed-forward networks:

\begin{equation}
	FFN(M_H) = max (M_H. W^{(1)} + b^{(1)} ) . W^{(2)}+b^{(2)}
\end{equation}

Higher-level shopping patterns emerge through repeated attention and feed-forward blocks:

\begin{equation}
	H^{b+1}=H(F^b) ,; F^{b+1}=FFN(H^{b+1})
\end{equation}

The decoder's additional cross-attention layer computes encoder outputs ($K$, $V$) with its own queries ($Q$) to connect historical patterns to current generation context. The final softmax layer generates probability distributions over categories for the next basket.

To stabilize generation and prevent overfitting, we employ skip connections, layer normalization, and dropout regularization \cite{vaswani2017Attention, kang2018Self}.

\subsection{Network training}
We train T-REX to generate next-basket predictions through teacher forcing. The decoder's inputs are right-shifted by one position, with causal masking ensuring predictions rely only on previous purchases. For each position, the model learns to predict item $v_t^u \in S_{dec}^u$ given prior purchases and the encoder context. The generation quality is optimized via cross-entropy loss:

\begin{equation}
	- \sum_{S \in S_{dec}} \sum_1^{|\mathcal{V}|} {v_{t+1} . \log (p(\hat{v}_{t+1}))}
\end{equation}

where $v_{t+1}$ represents the true next category (one-hot encoded) and $p(\hat{v}_{t+1})$ is the model's generated probability distribution over categories.

Our model architecture consists of a 6-layer encoder and 6-layer decoder, each with 6 attention heads and hidden dimension of 2048. The embedding dimension $d$ is set to 512, resulting in approximately 50.4M trainable parameters. While our dataset contains over 1MM shopping sessions from 100K customers, we considered approximately 900K sessions for training, requiring customers to have at least 3 shopping sessions in their history. This resulted in roughly 85K qualified customers for model training.

To stabilize training and improve generation quality, we employ Adam optimizer with learning rate $10^{-4}$ and weight decay 0.01, gradient clipping with max norm 0.5, batch size 32, and early stopping with 5-epoch patience.

\subsection{Inference}
During inference, T-REX generates probability distributions over all categories $v \in \mathcal{V}$, with top-k predictions forming the recommended basket. While the model supports dynamic basket generation through teacher forcing \cite{Kolen2001A} - updating predictions as customers add items - such real-time generation requires dedicated inference infrastructure and remains a future deployment target.

\section{Dataset}
The grocery purchase history dataset comprises over 1MM shopping sessions from 100K customers, with transactions chronologically ordered by date.

To balance efficiency and relevance, we map products to categories using Amazon's Grocery taxonomy. Starting with 35 high-level categories for initial validation, we later expanded to 280 finer-grained categories to increase recommendation specificity while maintaining computational efficiency.

This mapping addresses fundamental challenges in grocery recommendation: data sparsity at the product level and class imbalance from varying purchase frequencies. The category-level approach aligns with how customers plan their shopping, making the generated recommendations more intuitive.

For evaluation, we held out each customer's final basket as test data. The remaining histories were split 90/10 for training and validation. Our positional encoding scheme accommodates the full study timeframe, capturing both short-term shopping patterns and seasonal trends.

\section{Baseline and metrics}

Personal top frequency (P-Top) serves as our primary baseline. The choice of P-Top as our primary baseline is well-justified by recent findings in the literature. Li et al. demonstrated that in domains with significant repeat purchasing patterns, such as grocery shopping, P-Top provides remarkably robust performance, often matching or outperforming more sophisticated recommendation systems \cite{li2023A}. While simpler than recent neural approaches, P-Top's strong performance makes it a challenging baseline to surpass, particularly in the grocery domain where customer habits tend to be relatively stable. This aligns with our own experimental observations, where even advanced architectures (encoder-only and decoder-only transformers) struggled to consistently outperform P-Top.

For each customer $u$, P-Top generates recommendations by ranking categories based on their historical purchase frequency. Formally, it outputs $ PTop (S^u , k) = { v_j \in S^u |  j \leq k } $, where $v_j$ is the $j^{th}$ most frequently purchased category in the customer's history $S^u$. For example, if a customer most frequently buys dairy products, followed by produce and beverages, these categories would form their top-3 recommendations regardless of recent shopping patterns or seasonal trends.

We evaluate our generated recommendations using precision and recall at varying basket sizes ($k$):

\begin{equation}
	recall@k = \sum_{v \in S^u_{k+1}} \{1 |  v \in S_{t+1} ^u \} / |S_{k+1}^u |
\end{equation}

\begin{equation}
	precision@k=\sum_{v \in S_{k+1}^u} \{1 |  v \in s_{t+1}^u \} / k
\end{equation}

Beyond standard metrics, we introduce a rank-matching evaluation framework using matrix $[x_{i,j}]$, where entry $(i,j)$ indicates how often a category with true rank $i$ receives predicted rank $j$. Here, rank refers to the category's purchase frequency in the customer's history - for example, if dairy products are a customer's most frequently purchased category, it has rank 1, their second most frequent category has rank 2, and so on. This metric specifically evaluates how well our generated rankings align with customers' established shopping patterns, directly measuring the practical utility of recommendations for basket building.

\section{Results and discussion}

\begin{figure*}
	\centering
	\includegraphics[width=0.8\textwidth]{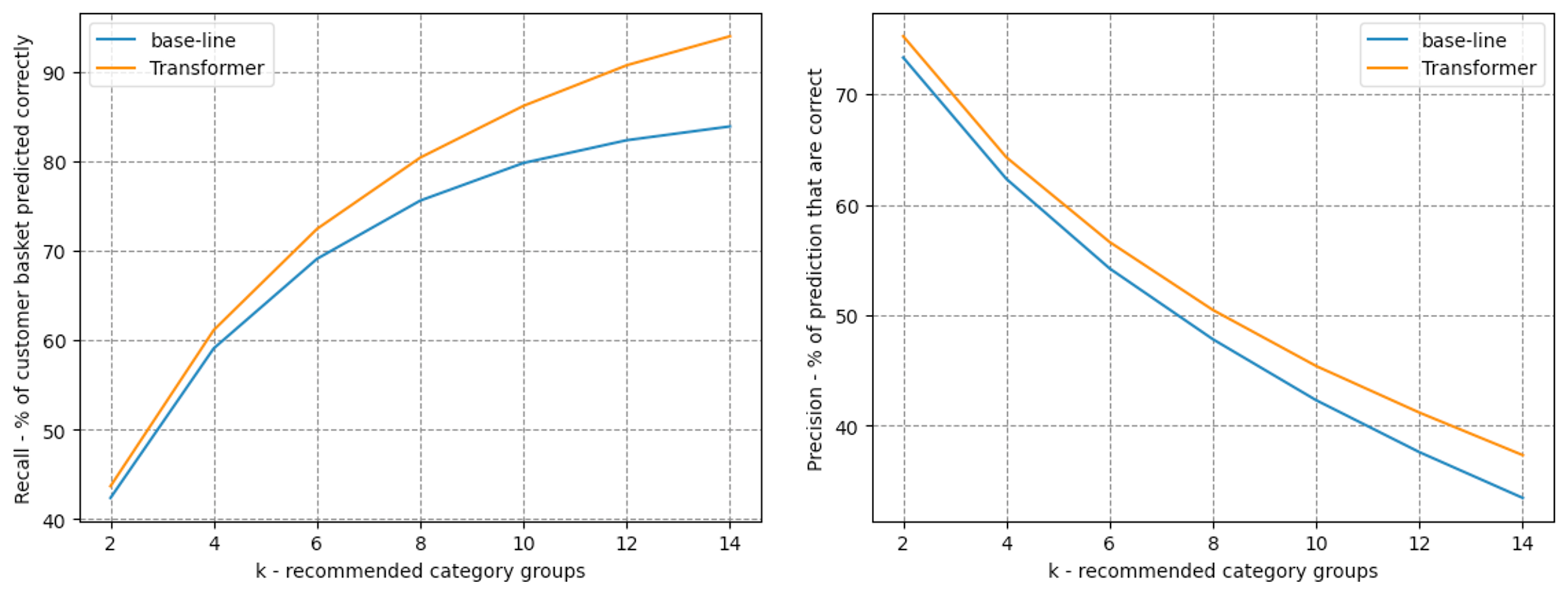}
	\caption{Comparison of average recall@k (left) and precision@k (right) between Transformer and P-Top models.}
\label{fig:prec_recall}
\end{figure*}

\begin{figure*}
	\centering
	\includegraphics[width=0.8\textwidth]{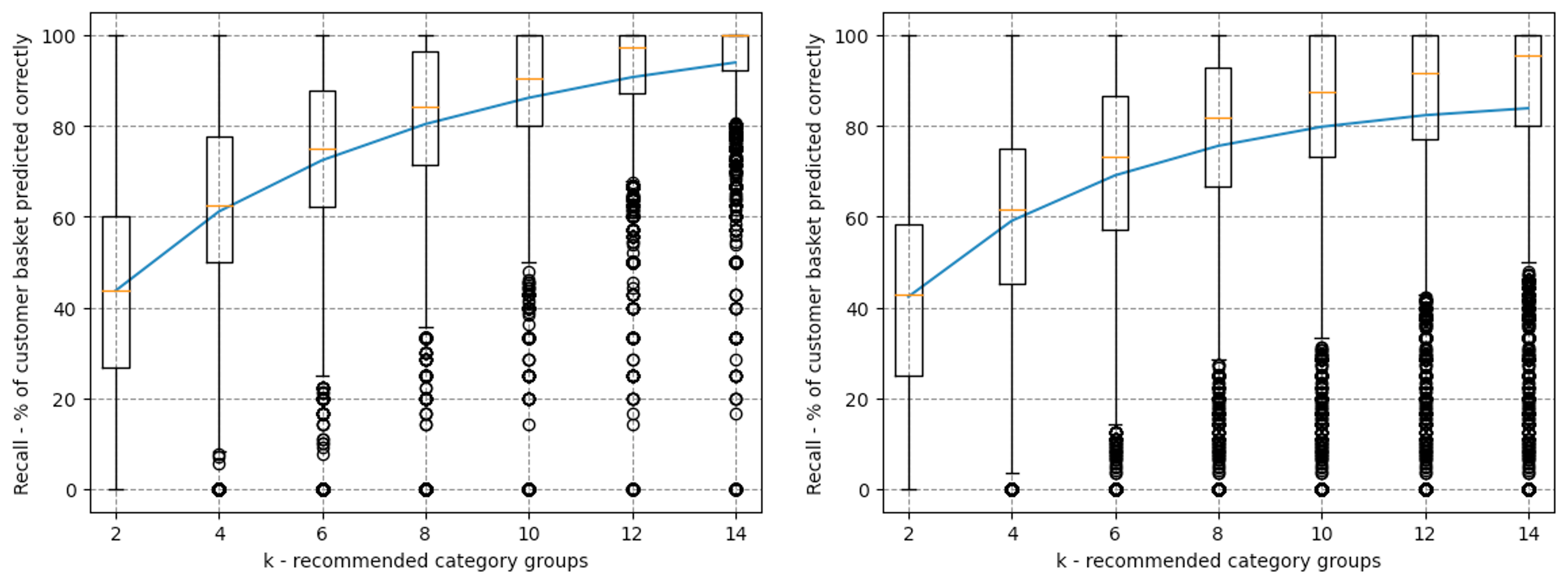}
	\caption{Variance analysis of recall@k for transformer (left) and P-Top baseline (right). Box-plots present the distribution of Recall at each value of ‘k’. The bottom of each box is the 25 percentile, the orange bar is the median, the top of the box is the 75 percentile and circles are the outliers.}
\label{fig:percent}
\end{figure*}

Figure \ref{fig:prec_recall} compares average precision@k and recall@k (k=2-14) between T-REX and P-Top baseline. Our model's generated recommendations outperform the baseline across all values of $k$. Figure \ref{fig:percent}, presents a variance analysis of recall@k through box-plots, illustrating the distribution's quartiles (25th percentile, median, and 75th percentile) across the aggregate results.

The median customer retrieved over 90\% of their basket items from T-REX's top 10 generated recommendations. Beyond higher average recall, T-REX demonstrates more consistent generation quality, evidenced by shorter whiskers and closer outliers in the box plots. This consistency indicates that the model learns generalizable shopping patterns rather than simply memorizing frequency-based rules.

\begin{figure*}
	\centering
	\includegraphics[width=0.8\textwidth]{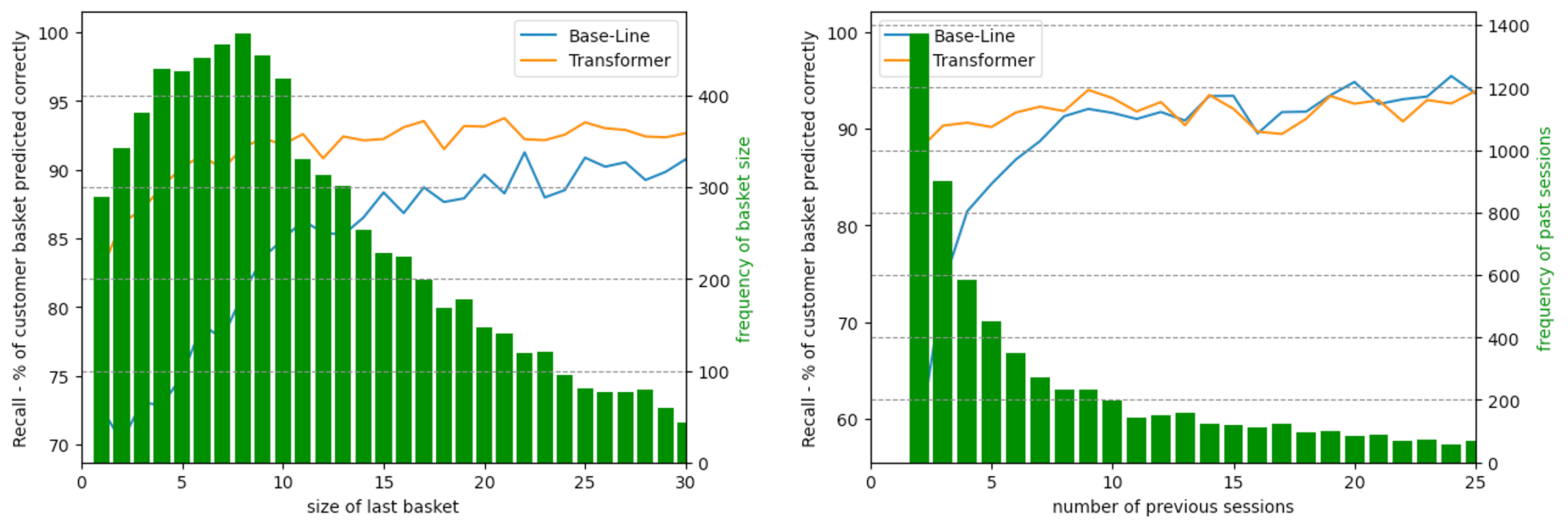}
	\caption{Breakdown of recall@10  for the number of previous sessions (right) as well as the length of the last basket (left). Results are from 10,000 sub-sample of the test dataset.}
\label{fig:sessions}
\end{figure*}

To \textit{protect customer privacy}, we present aggregated performance metrics across customer segments. We analyzed two key factors affecting model performance (Figure \ref{fig:sessions}):

\textbf{Customer Tenure}: For established customers (8-10+ previous sessions), both approaches show comparable performance. This convergence likely reflects several factors. First, long-term customers often develop stable shopping patterns across a limited set of categories, making frequency-based predictions increasingly effective. Additionally, many customers may concentrate their online grocery shopping within specific category subsets while purchasing other categories through alternative channels. While some customers may have fewer than k historical items for recall@k evaluation, this limitation affects both models equally - the baseline cannot generate enough recommendations while T-REX's additional predictions are counted as false positives, ensuring fair comparison. The baseline's performance improvement with tenure thus reflects the predictability of established shopping patterns.

\textbf{Basket Size}: Both approaches face challenges with small baskets ($<$15 items), but T-REX maintains a 10\% advantage, increasing to 30\% for very small baskets ($<$5 items). This demonstrates our model's ability to generate accurate recommendations even with limited context. While the baseline relies solely on personal purchase history, which may not provide enough signal for customers with limited category exposure, T-REX can leverage patterns learned from the broader customer base. The performance gap narrows for larger baskets ($>$30 items), where frequency-based patterns become more reliable.

\begin{figure*}
	\centering
	\includegraphics[width=0.8\textwidth]{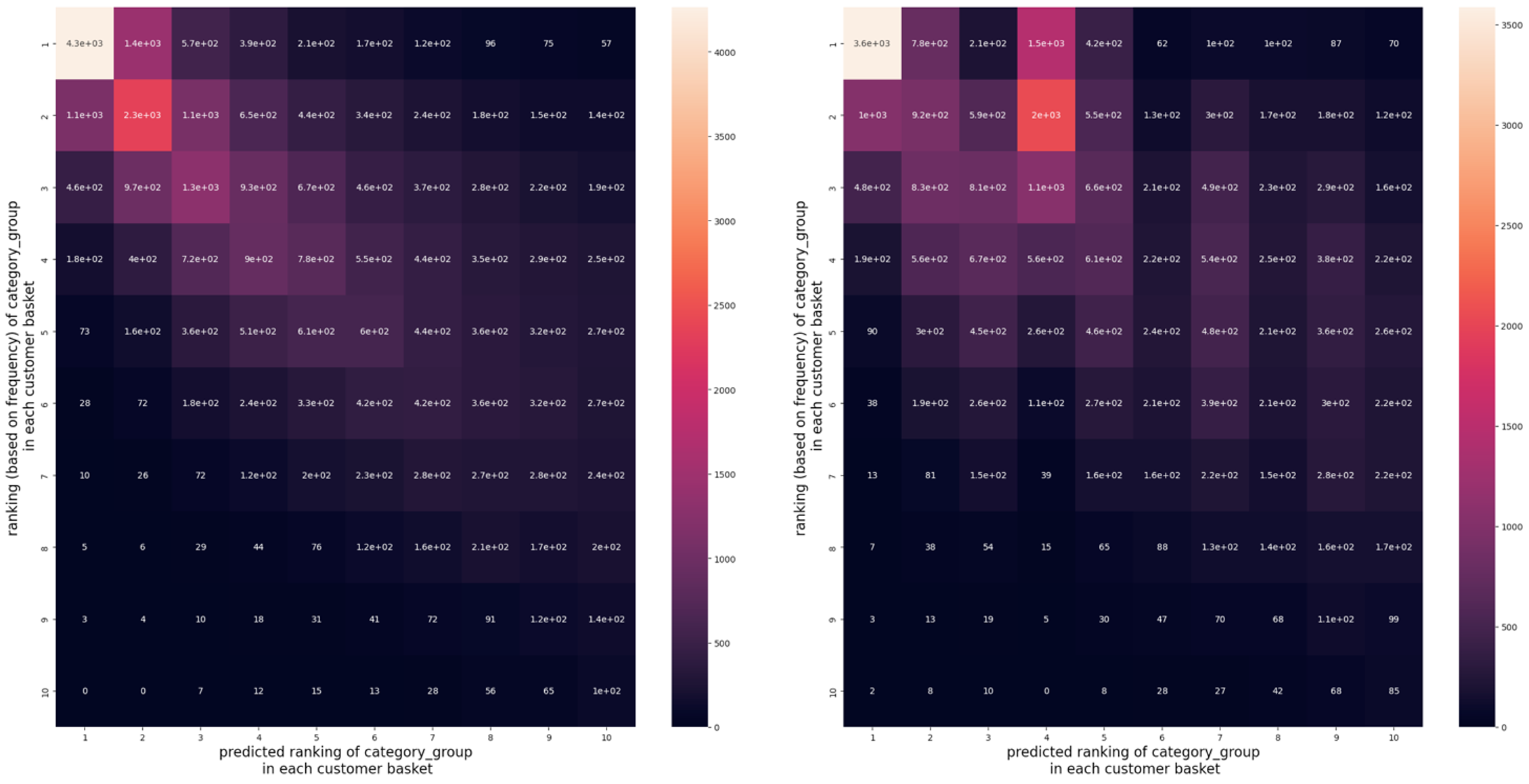}
	\caption{The proposed rank-matching metric, comparing the actual ranking of categories in the customers' last basket with the predicted rankings from the transformer model (left) and rankings from the P-Top baseline (right).}
\label{fig:rank}
\end{figure*}

The rank-matching analysis (Figure \ref{fig:rank}) provides deeper insights into generation quality. We compare category rankings based on actual purchase volumes against generated recommendation rankings. The resulting confusion matrices visualize ranking alignment - each cell $(i,j)$ shows how often a category with true rank $i$ received predicted rank $j$. T-REX achieves superior ranking generation with 30\% exact matches (vs 27\% for P-Top) and 60\% within one rank (vs 48\% for P-Top). This improved alignment translates directly to more efficient basket building, as customers find their most important categories earlier in the recommendation sequence.

\section{Production deployment and A/B testing}
We validated T-REX through production A/B testing, focusing on Consumables recommendations. For production deployment, we expanded our generation capabilities to 280 finer-grained categories while maintaining model efficiency. The system demonstrated exceptional performance, achieving the highest widget clickthrough rates among comparable recommendation widgets and delivering significant improvements in order product sales, composite sales, and gross contribution profit.  Notably, the sales lift from T-REX was approximately 23\% higher than typical (Amazon-wide) recommendation systems.

Our A/B testing methodology employed customer-level randomization across various placement locations, including pre- and post-checkout pages. We chose customer-level over session-level randomization due to its higher statistical power. To quantify the average treatment effect (ATE), we applied empirical Bayes methods following the framework detailed in \cite{masoero2023leveraging}. Additionally, we analyzed long-term downstream effects using double-machine learning (DML) causal methods \cite{chernozhukov2018double}, which revealed sustained positive incremental effects of our recommendations over time.

\section{Conclusions}
We presented T-REX, a transformer-based architecture for generating personalized grocery basket recommendations. Our key innovations extend beyond standard transformer architectures through three main components: dynamic sequence splitting for improved generalization across varying customer histories, adaptive positional encoding to capture irregular temporal patterns in grocery shopping, and category-level modeling that balances computational efficiency with intuitive recommendations.

Experimental results demonstrate significant improvements over traditional frequency-based approaches. T-REX shows particularly strong performance for new customers, with substantial improvements in recall metrics and ranking accuracy. The system's ability to generate more relevant recommendations translates directly to improved basket building efficiency for customers, as validated through production deployment.

While T-REX demonstrates strong performance in production, several opportunities for future research remain. These include comprehensive comparisons with other transformer-based recommenders (e.g., SASRec, BERT4Rec) and ablation studies of key components like dynamic sequence splitting and adaptive positional encoding. Additionally, future work should investigate potential performance variations across different customer segments and shopping patterns to further optimize the recommendation experience.

\bibliography{aaai2026}
\bibliographystyle{aaai2026}

\end{document}